\documentclass{svproc}
\usepackage{graphicx}
\usepackage{upgreek}
\usepackage{multirow}
\usepackage{color}
\usepackage[mathlines]{lineno}% Enable numbering of text and display math

\usepackage[colorlinks=true, pdfstartview=FitV, linkcolor=blue, citecolor=black, plainpages=false, pdfpagelabels=true, urlcolor=blue]{hyperref}

\begin{document}
% \linenumbers
\mainmatter              % start of a contribution
\title{Recent Heavy-Flavor Results from STAR}
\titlerunning{Recent Heavy-Flavor Results from STAR}  % abbreviated title (for running head)
\author{Guannan Xie (for the STAR Collaboration)}
\authorrunning{Guannan Xie} % abbreviated author list (for running head)
\tocauthor{Guannan Xie}
\institute{Lawrence Berkeley National Laboratory, Berkeley CA 94706, USA 
\email{xieguannanpp@gmail.com}}

\maketitle              % typeset the title of the contribution

\begin{abstract}
  In these proceedings, we report on the production of various open heavy-flavor hadrons and quarkonia in Au+Au collisions at $\sqrt{s_{\rm{NN}}}\ =\ 200\ \textup{GeV}$ from the STAR experiment.
% We would like to encourage you to list your keywords within
% the abstract section using the \keywords{...} command.
\end{abstract}
\section{Introduction}

Due to the intrinsic large mass (charm and bottom), measurements of heavy-flavor production (open heavy-flavor hadrons and quarkonia) are an important tool for studying the properties of the Quark-Gluon Plasma (QGP) formed in relativistic heavy-ion collisions. The modification of their distributions in transverse momentum ($p_{T}$) due to energy loss and in azimuth due to anisotropic flows is sensitive to heavy-quark dynamics in the partonic QGP phase~\cite{intro_1,intro_2}.

In these proceedings we present measurements of the $D^0$ nuclear modification factors and elliptic flow in Au+Au collisions from STAR, and compare to similar measurements for light-flavor hadrons. The $\Lambda_{c}^{\pm}$ and $D_{s}^{\pm}$ production are presented to study the coalescence mechanism for charm-quark hadronization. The measurements of open bottom production through the reconstruction of their displaced decay daughters (B $\rightarrow J/\psi,D^{0},e$) are performed to test the mass dependence of parton-medium interactions in the QGP. The strong $J/\psi$ suppression in heavy-ion collisions has a complicated interpretation as not only color-screening, but also the cold nuclear matter (CNM) effects and the regeneration mechanism play a role. $\Upsilon$ measurements are a cleaner probe of the color-screening effect at RHIC energies and the suppression pattern of different bottomonium states will help to constrain the temperature of the medium. The $J/\psi$ measurements as well as the $\Upsilon$ ones are also presented in these proceedings.

\section{Nuclear modification factors for $D^0$}

\begin{figure}[htbp]
\hspace{+1.0cm}
\begin{minipage}[b]{0.35\linewidth}%0.5
\begin{center}
  \includegraphics[width=\textwidth]{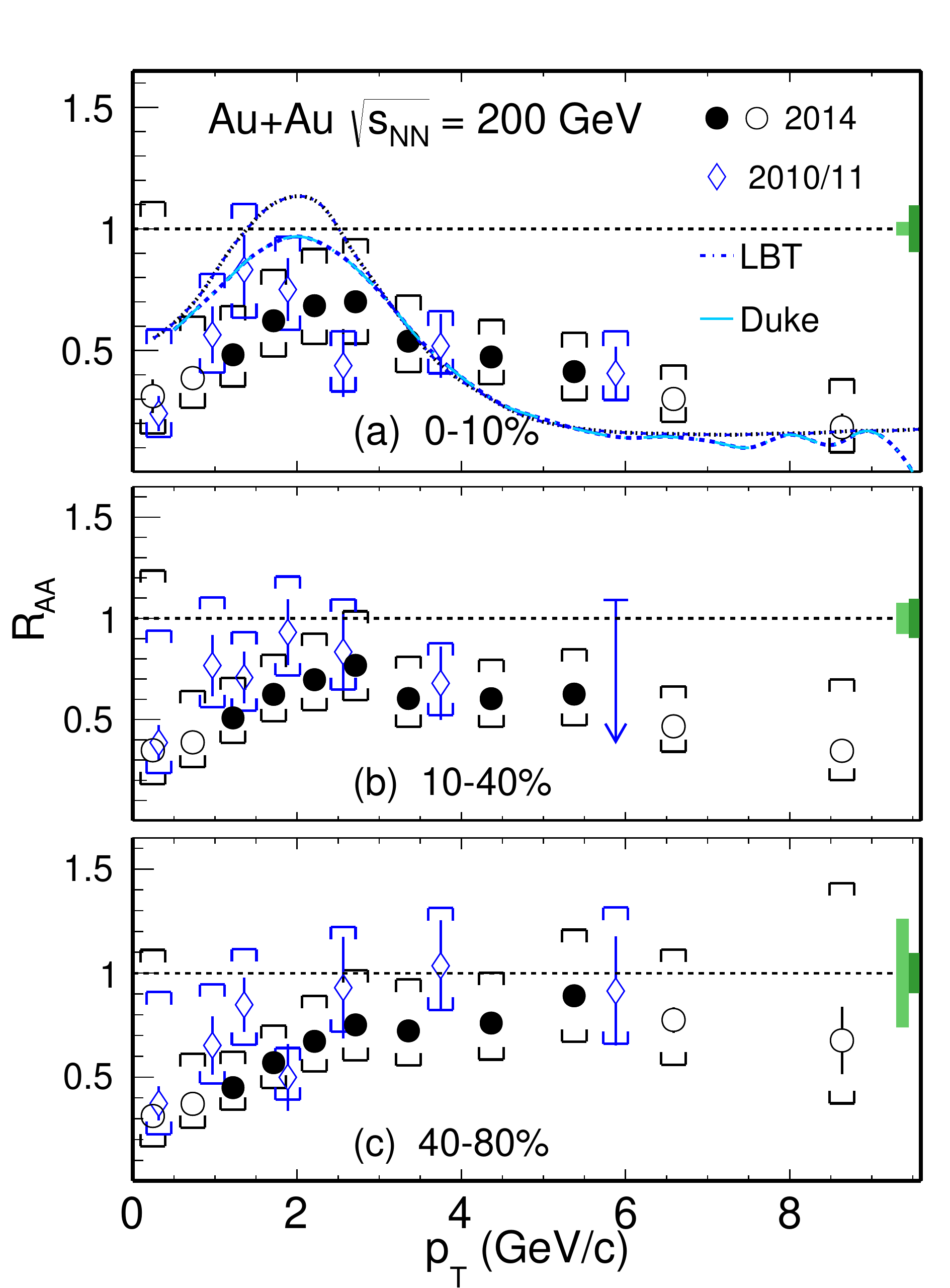}
\end{center}
\end{minipage}
\hspace{+1.5cm}
\begin{minipage}[b]{0.35\linewidth}%0.57
\begin{center}
  \includegraphics[width=\textwidth]{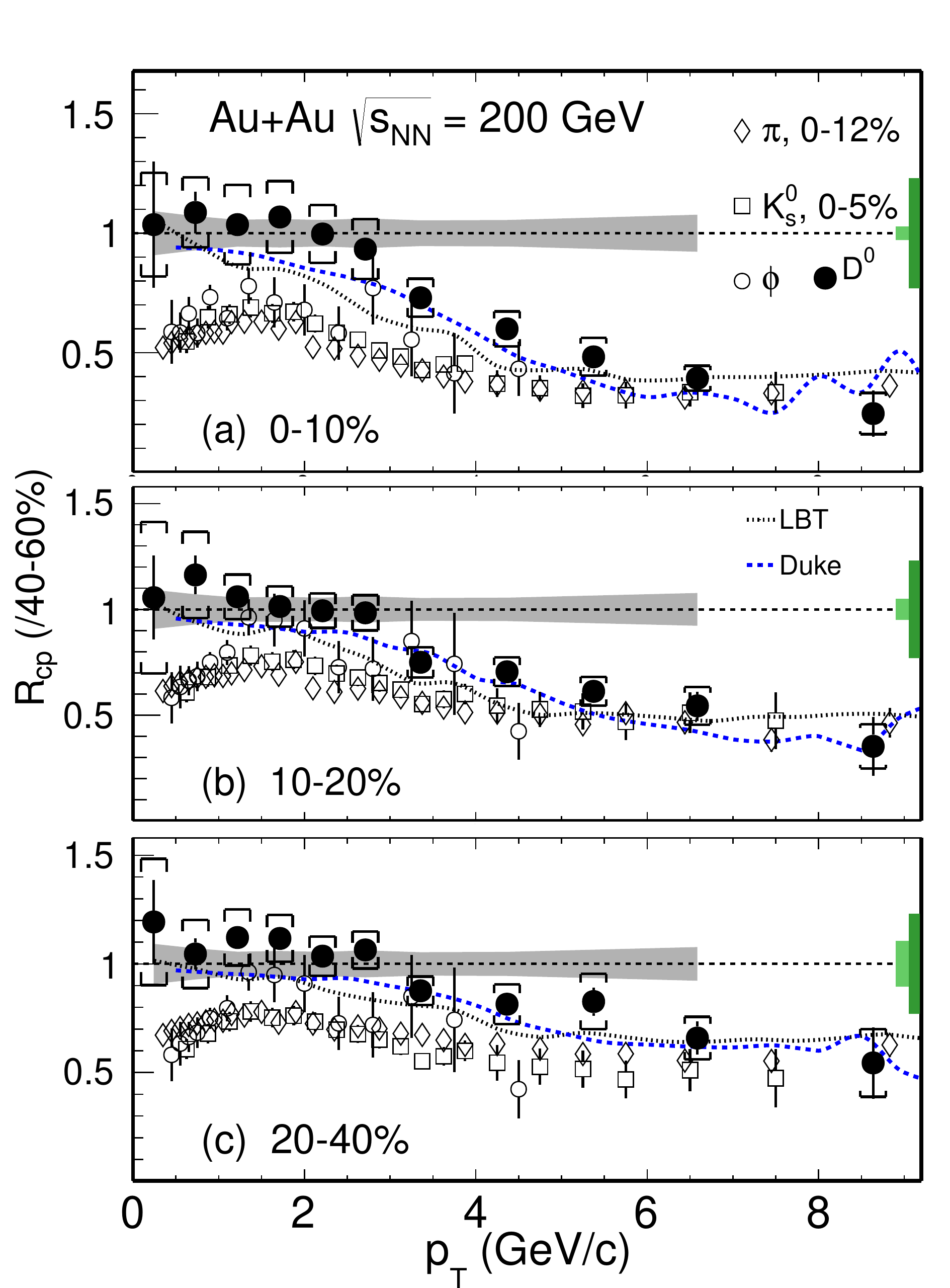}
\end{center}
\end{minipage}
\caption{ (Left) $D^{0}$ $R_{\rm AA}$ in Au+Au collisions at $\sqrt{s_{\rm{NN}}}$ = 200 GeV for different centrality bins. (Right) $D^{0}$ $R_{\rm CP}$ with the 40--60\% spectrum as the reference.}
\label{fig:D0_RAA_Rcp}
\end{figure}

Figure~\ref{fig:D0_RAA_Rcp} left panel shows the $D^0$ $R_{\rm AA}$, which is the yield ratio between Au+Au and p+p scaled by the number of binary collisions~\cite{Star_D0_HFT}. From low to intermediate $p_{T}$ region, the $D^0$ $R_{\rm AA}$ shows a characteristic structure which is qualitatively consistent with the expectation from model predictions in which charm quarks gain sizable collective motion during the medium evolution. In order to take advantage of the precision of the Au+Au spectra and avoid the large uncertainties from the p+p baseline, we construct the $R_{\rm CP}$ which is the yield ratio between central and peripheral Au+Au collisions. The right panel shows the $D^0$ $R_{\rm CP}$ for different centralities as a function of $p_{T}$ with the 40--60\% centrality spectrum as the reference. The measured $D^0$ $R_{\rm CP}$ in central 0--10\% collisions shows a significant suppression at $p_{T}>$ 5\,GeV/$c$. The suppression level is similar to that of light-flavor hadrons and strange mesons and the suppression gradually decreases from central to mid-central and peripheral collisions, similarly as $R_{AA}$. The $D^0$ $R_{\rm CP}$ for $p_{T}$\,$<$\,4\,GeV/$c$ does not show a modification with centrality, in contrast to light-flavor hadrons. Calculations from the Duke group and the Linearized Boltzmann Transport (LBT) model match the data well~\cite{duke,lbt}, while the improved precision of the new measurements is expected to further help constrain the theoretical model calculations.

\section{$D_s/D^0$, $\Lambda_c/D^0$ yield ratios}

\begin{figure}[htbp]
\hspace{+0.3cm}
\begin{minipage}[b]{0.287\linewidth}%0.5
\begin{center}
\includegraphics[width=\textwidth]{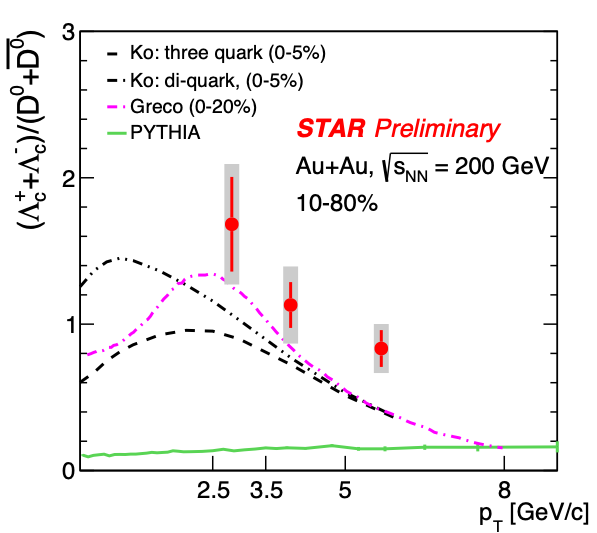}
\end{center}
\end{minipage}
\hspace{+0.0cm}
\begin{minipage}[b]{0.275\linewidth}%0.5
\begin{center}
\includegraphics[width=\textwidth]{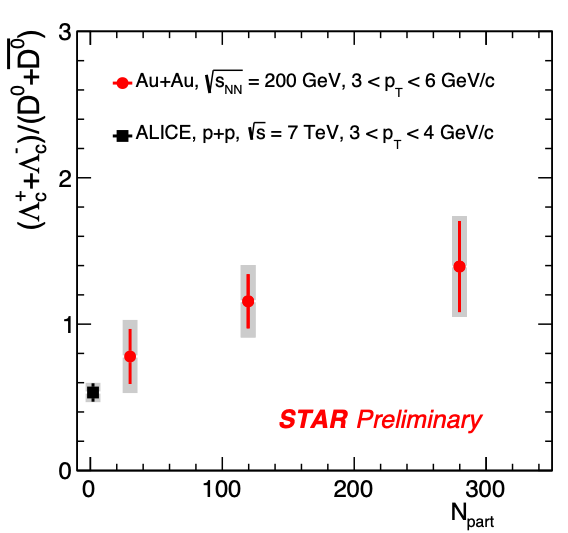}
\end{center}
\end{minipage}
\hspace{+0.0cm}
\begin{minipage}[b]{0.285\linewidth}%0.5
\begin{center}
\includegraphics[width=\textwidth]{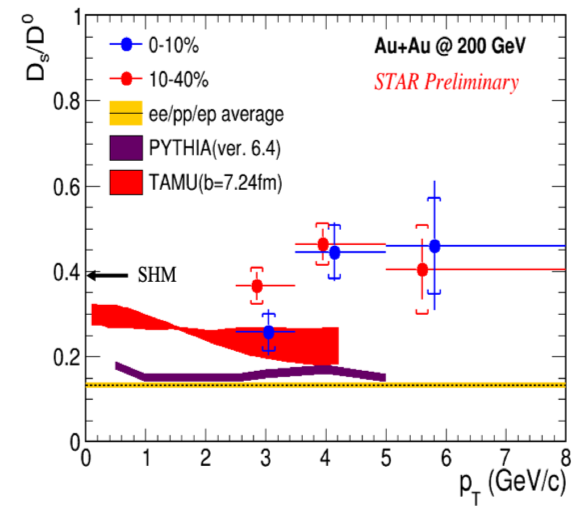}
\end{center}
\end{minipage}
  \caption{ (Left) $\Lambda_c/D^0$ ratio as a function of $p_T$ for the 10--80\% centrality class. (Middle) $\Lambda_c/D^0$ ratio as a function of $N_{\rm part}$ in $3 < p_{T} < 6$\,GeV/$c$. (Right) $D_s/D^0$ ratio as a function of $p_T$ for the 0--10\% and 10--40\% centralities.}
\label{fig:Lc_pt_cent}
\end{figure}

Figure~\ref{fig:Lc_pt_cent} left panel shows the $\Lambda_c/D^0$ yield ratio as a function of $p_T$ for the \mbox{10--80\%} centrality class. The values show a significant enhancement compared to the calculations from PYTHIA.  The model calculations which include coalescence hadronization of charm quarks can qualitatively reproduce the $p_T$ dependence~\cite{shm,Ko,greco}. However, one needs measurements at low $p_T$ to further differentiate between different models. The middle panel shows the measured $\Lambda_c/D^0$ ratio as a function of $N_{\rm part}$ in $3 < p_{T} < 6$\,GeV/$c$. There is a clear increasing trend towards more central collisions while the value in the peripheral collisions is comparable with the measurement in $p$+$p$ collisions at $\sqrt{s_{\rm{NN}}}$ = 7 TeV from ALICE~\cite{aliceLc}. The right panel shows the $D_s/D^0$ ratio for two centrality classes. There is a strong enhancement compared to the PYTHIA fragmentation with no significant centrality dependence~\cite{STAR_Ds}.

Besides the $D^0$, $D_s$ and $\Lambda_c^{\pm}$, STAR also has performed measurements of $D^{\pm}$ in Au+Au collisions at $\sqrt{s_{\rm{NN}}}$ = 200 GeV. With these various charmed hadron measurements, the total charm quark cross section per binary nucleon-nucleon collision was obtained as listed in Table 1. The total $c\overline{c}$ cross section per binary nucleon-nucleon collision in Au+Au collisions is consistent with that in $p$+$p$ within uncertainties. However, as demonstrated by the $\Lambda_c/D^0$ and $D_s/D^0$ yield ratios, the charm hadrochemistry is modified in heavy-ion collisions compared to $p$+$p$ collisions.

\begin{table*}
  \small
\centering{
\caption{Total charm cross-section per binary nucleon-nucleon collision at midrapidity in Au+Au and $p$+$p$ collisions at 200 GeV.}
\begin{tabular}{c|c|c} \hline \hline
  \multicolumn{2}{c|} {Charm Hadron}   & Cross Section d$\sigma$/dy($\upmu$b) \\ \hline
  \multirow{4}{*}{Au+Au } & \hspace{0.8cm} $D^0$ \hspace{0.8cm}  & 41 $\pm$ 1 (stat) $\pm$ 5 (sys) \\ \cline{2-3} 
  & \hspace{0.8cm} $D^+$ \hspace{0.8cm} & 18 $\pm$ 1 (stat) $\pm$ 3 (sys)\\ \cline{2-3}
  \multirow{2}{*}{(10-40\%)} & \hspace{0.8cm} $D_s^+$ \hspace{0.8cm} & 15 $\pm$ 1 (stat) $\pm$ 5 (sys)\\ \cline{2-3}
  & \hspace{0.8cm} $\Lambda_c^+$ \hspace{0.8cm} & 78 $\pm$ 13 (stat) $\pm$ 28 (sys)\\ \cline{2-3}
  & \hspace{0.8cm} total $c\overline{c}$ \hspace{0.8cm} & 152 $\pm$ 13 (stat) $\pm$ 29 (sys)\\ \hline 
  $p$+$p$ & \hspace{0.8cm} total $c\overline{c}$ \hspace{0.8cm} & 130 $\pm$ 30 (stat) $\pm$ 26 (sys)\\ \hline \hline
\end{tabular}
}
\label{table:crossX}
\end{table*}

\section{$D^0$ elliptic flow ($v_2$) and directed flow ($v_1$)}

\begin{figure}[htbp]
\hspace{+0.01cm}
\begin{minipage}[b]{0.295\linewidth}%0.5
\begin{center}
  \includegraphics[width=\textwidth]{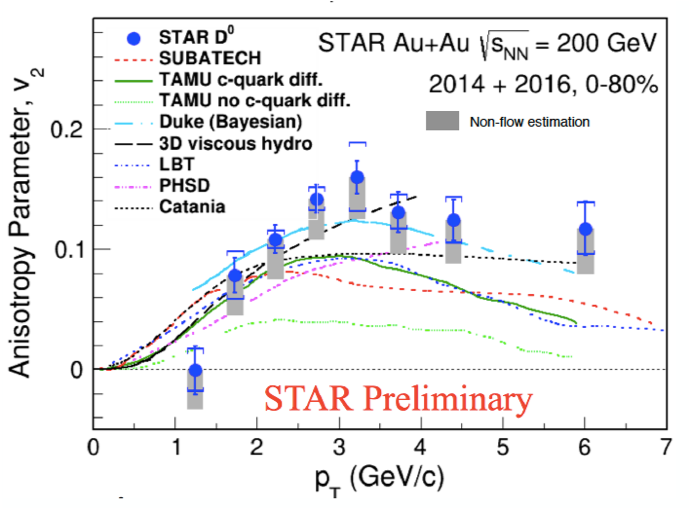}
\end{center}
\end{minipage}
\hspace{+0.01cm}
\begin{minipage}[b]{0.315\linewidth}%0.57
\begin{center}
  \includegraphics[width=\textwidth]{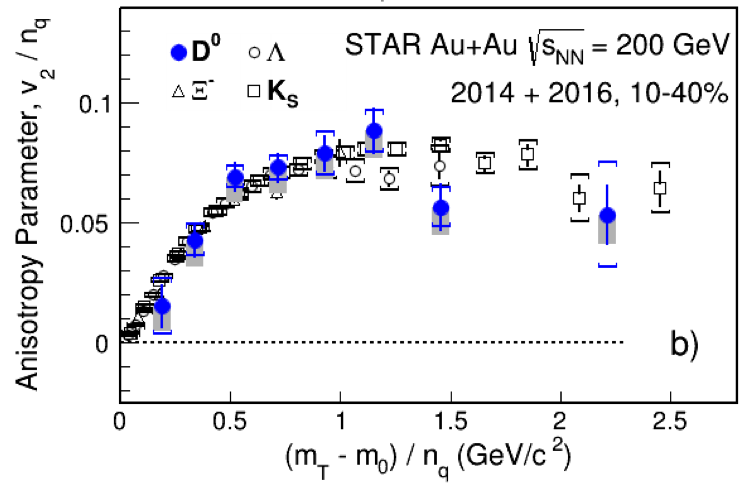}
\end{center}
\end{minipage}
% \put(-110, 22) {\footnotesize \color{red} STAR Preliminary}
\hspace{+0.01cm}
\begin{minipage}[b]{0.336\linewidth}%0.57
\begin{center}
  \includegraphics[width=\textwidth]{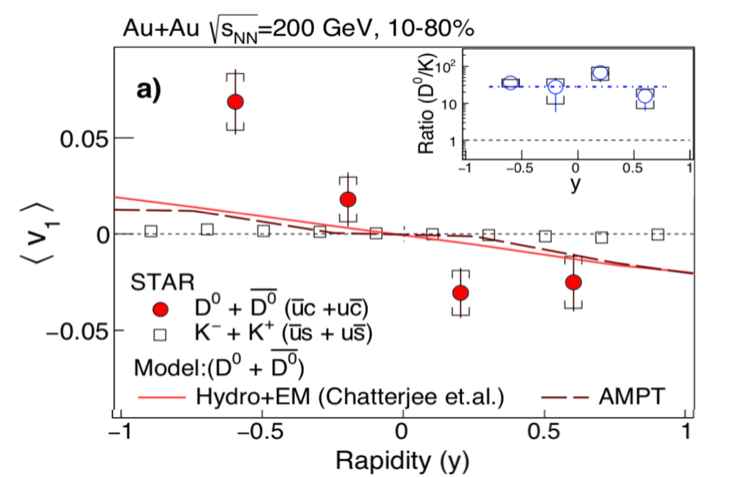}
\end{center}
\end{minipage}
  \caption{$D^0$ elliptic flow $v_2$ vs $p_T$ (Left) and the test of NCQ scaling (Middle) for $D^0$ and light-flavor hadrons. (Right) $D^0$ directed flow $v_1$ vs rapidity.}
\label{fig:D0_v2}
\end{figure}

Figure~\ref{fig:D0_v2} left panel shows STAR results showing a large non-zero $v_2$ for $D^0$ mesons~\cite{STAR_d0_v2}, which clearly follows the Number of Constituent Quarks (NCQ) scaling similarly as light-flavor hadrons below $p_T$ of 1 GeV/$c$ as shown in the middle panel. This suggests that charm quarks gain significant flow through interactions with the medium. The $v_2$ is compared to various model calculations and in particular the 3D viscous hydrodynamic model calculation can reproduce the results for $p_T < 4$ GeV/c. The other transport models with charm quark diffusion in the medium are consistent with the data when incorporating a diffusion coefficient (2$\pi$$T$$D_s$) in the range of $2\sim5$ around $T_c$~\cite{xin_review}.

The $D$-meson directed flow $v_1$ is expected to be sensitive to the initial tilt of the bulk medium, while the difference between $D^0$ and $\overline{D^{0}}$ is suggested to be sensitive to the initial electromagnetic field. Figure~\ref{fig:D0_v2} right panel shows the first observation of a non-zero $D$-meson $v_1$ slope which is much larger than that of kaons. The $v_1$ values measured separately for $D^0$ and $\overline{D^0}$ are consistent within uncertainties. Future measurements with improved precision are needed to investigate the potential influence of the electromagnetic field on the $v_1$ values~\cite{STAR_d0_v1}.

\section{Measurements of $R_{AA}$ for $B$-decayed $J/\psi$, $D^0$ and $e$}

\begin{figure}[htbp]
\hspace{+0.5cm}
\begin{center}
  \includegraphics[width=0.86\textwidth]{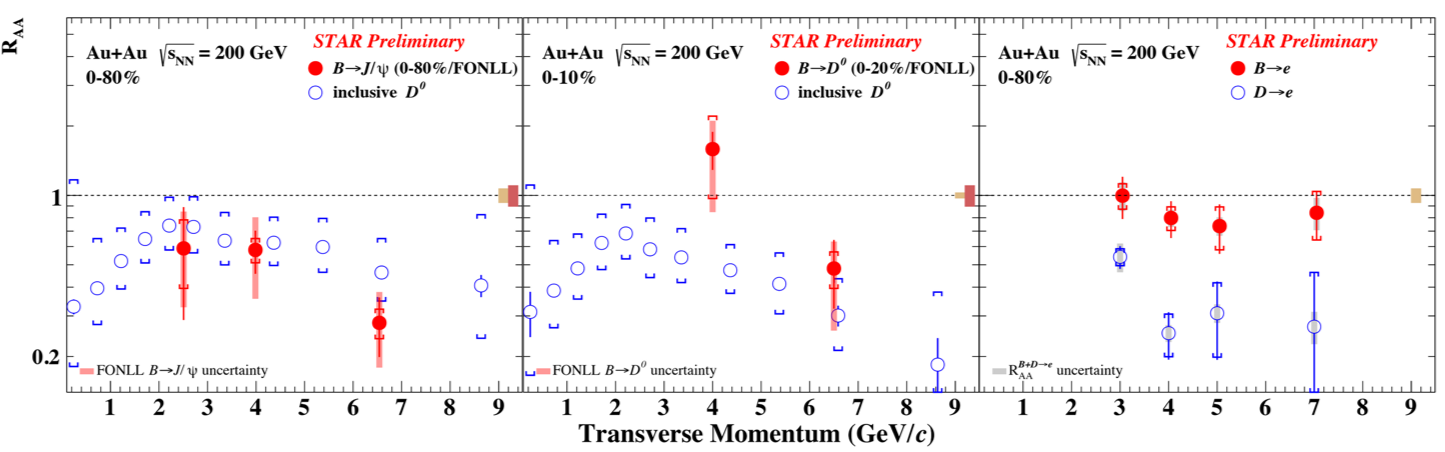}
\end{center}
  \caption{$R_{AA}$ of different daughter particles from decays of B-hadrons including \mbox{$B$$\rightarrow$ $J/\psi$}, $B$$\rightarrow$$D^0$ and $B$$\rightarrow$$e$.}
\label{fig:b_raa}
\end{figure}

The STAR Heavy Flavor Tracker (HFT) provides the capability of using the impact parameter method to distinguish the daughter particles from decays of bottom hadrons. Figure~\ref{fig:b_raa} shows the $R_{AA}$ of $B$$\rightarrow$$J/\psi$, $D^0$ and $e$. Strong suppressions for $B$$\rightarrow$$J/\psi$ and $B$$\rightarrow$$D^0$ at high $p_T$ are observed. The production of $B$$\rightarrow$$e$ is less suppressed than that of $D$$\rightarrow$$e$ with a significance level of about 2$\sigma$, which is consistent with the expectation of mass hierarchy of parton energy loss~\cite{shenghui_B_panic}.

\section{Measurements of $J/\psi$ productions in Au+Au collisions}

\begin{figure}[htbp]
\hspace{+1.0cm}
\begin{minipage}[b]{0.375\linewidth}%0.5
\begin{center}
  \includegraphics[width=\textwidth]{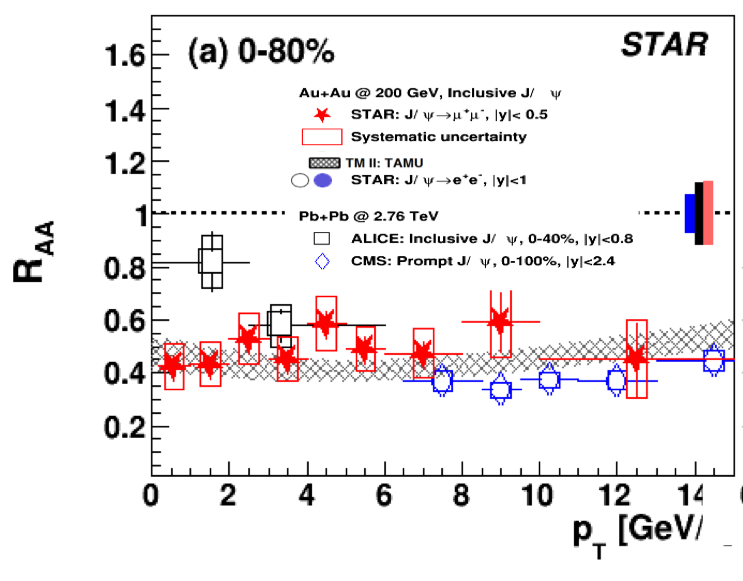}
\end{center}
\end{minipage}
\hspace{+1.0cm}
\begin{minipage}[b]{0.365\linewidth}%0.57
% \vspace{+1.2cm}
\begin{center}
  \includegraphics[width=\textwidth]{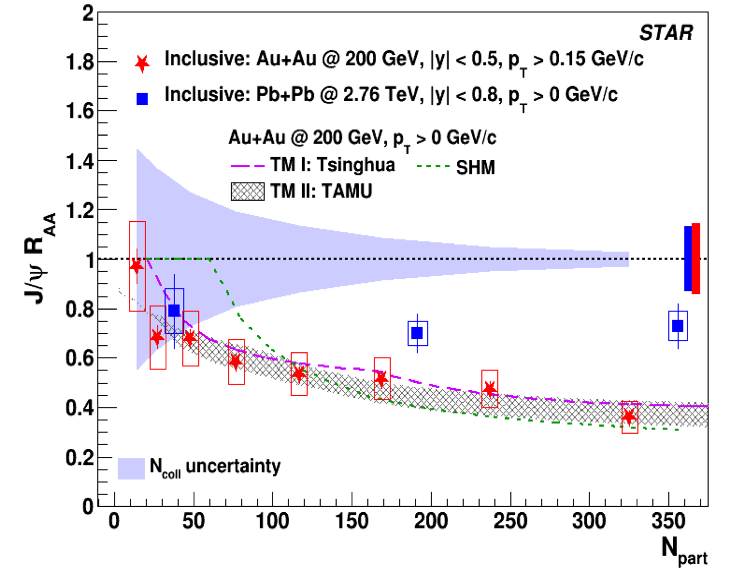}
\end{center}
\end{minipage}
  \caption{(Left) $J/\psi$ $R_{AA}$ as a function of $p_T$ in Au+Au collisions at $\sqrt{s_{\rm{NN}}}$ = 200 GeV. (Right) $J/\psi$ $R_{AA}$ as a function of $N_{part}$ in Au+Au collisions, compared to that in Pb+Pb collisions at $\sqrt{s_{\rm{NN}}}$ = 2.76 TeV and model calculations.}
\label{fig:jpsi_raa}
\end{figure}

Figure~\ref{fig:jpsi_raa} shows the $J/\psi$ $R_{AA}$ reconstructed through the di-muon channel using the Muon Telescope Detector (MTD) as a function of $p_T$ in Au+Au collisions~\cite{STAR_jpsi_raa}. As can be seen the $J/\psi$ production is suppressed across the whole $p_T$ range. The suppression at low $p_T$ is likely due to the combination of the cold nuclear matter effects, the regeneration and the dissociation in the QGP. With increasing $p_T$ the CNM effects are expected to diminish. The relative contribution from the b-hadron decays increases with $p_T$, and the suppression level of $J/\psi$ originating from these decays is expected to be smaller than that of the prompt $J/\psi$. The centrality dependence of the $J/\psi$ suppression is shown in the right panel. The $R_{AA}$ decreases from peripheral to central collisions. Comparing the Au+Au measurements at $\sqrt{s_{\rm{NN}}}$ = 200 GeV to the Pb+Pb measurements at $\sqrt{s_{\rm{NN}}}$ = 2.76 TeV from the LHC~\cite{alice_jpsi_raa,cms_jpsi_raa}, the STAR result shows more suppression in central and semi-central collisions, which is likely due to a smaller contribution from regeneration caused by the lower charm production cross-section at the RHIC energy. Models taking into account dissociation and regeneration can reasonably describe the data~\cite{jps_model1,jps_model2,jps_model3}.

\section{Measurements of $\Upsilon$ productions in Au+Au collisions}

\begin{figure}[htbp]
\hspace{+0.8cm}
\begin{minipage}[b]{0.40\linewidth}%0.5
\begin{center}
  \includegraphics[width=\textwidth]{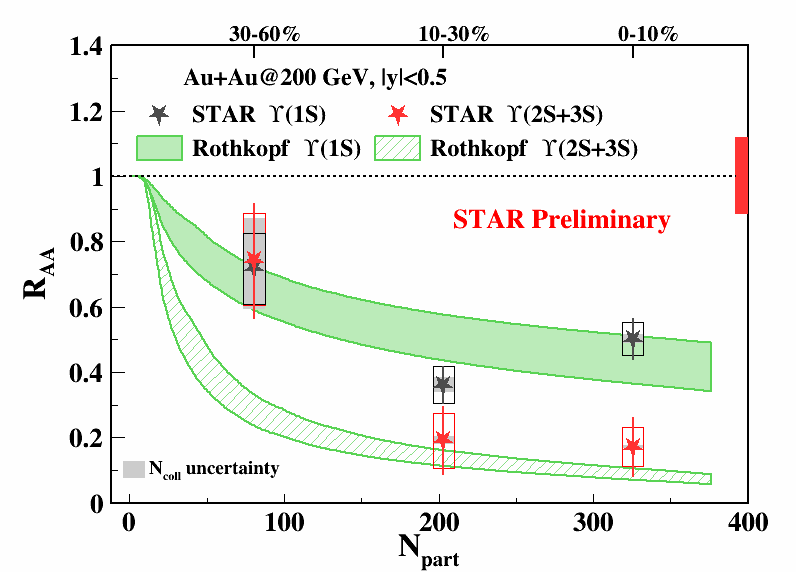}
\end{center}
\end{minipage}
\hspace{+0.8cm}
\begin{minipage}[b]{0.40\linewidth}%0.57
\begin{center}
  \includegraphics[width=\textwidth]{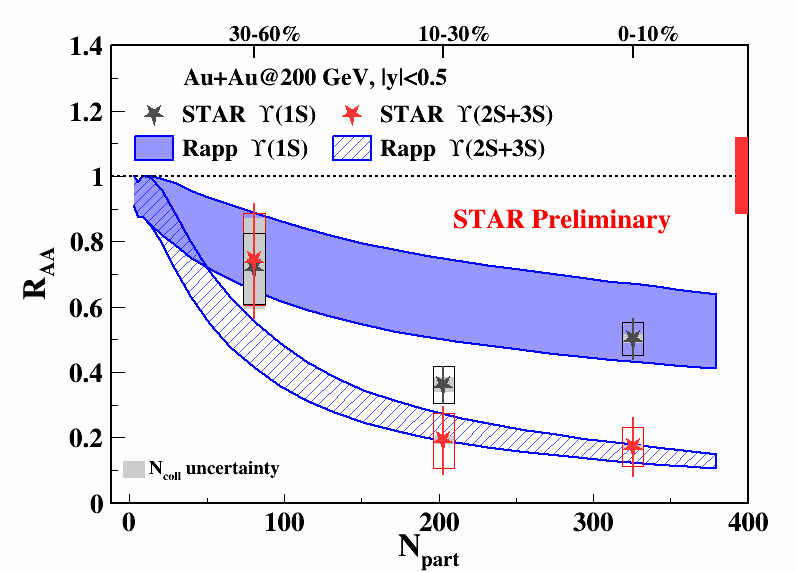}
\end{center}
\end{minipage}
  \caption{$\Upsilon$(1S) and $\Upsilon$(2S+3S) $R_{AA}$ as a function of $N_{part}$ in Au+Au collisions, compared to two model calculations.}
\label{fig:Upsilon_raa}
\end{figure}

Figure~\ref{fig:Upsilon_raa} shows the $\Upsilon$(1S) and $\Upsilon$(2S+3S) $R_{AA}$ as a function of $N_{part}$ in Au+Au collisions from the combined dielectron and dimuon results. The $R_{AA}$ shows a decreasing trend from peripheral to central collisions for both $\Upsilon$ $R_{AA}$, while the $\Upsilon$(2S+3S) are more suppressed than $\Upsilon$(1S) in the most central collisions. This is consistent with the ``sequential melting'' expectation. The data are also compared with two model calculations. In the Rothkopf model~\cite{Upsilon_model1} the $\Upsilon$ behavior in the QGP medium is described using a complex potential from lattice QCD calculations and there are no CNM or regeneration effects. While in the Rapp model~\cite{Upsilon_model2}, both CNM and regeneration effects are taken into account. These two models can describe well the measurements for the $\Upsilon$(1S) and $\Upsilon(2S+3S)$ in mid-central and central collisions.

\section{Summary}

We have presented the recent measurements of various open heavy-flavor hadrons in Au+Au collisions at $\sqrt{s_{\rm{NN}}}$ = 200 GeV utilizing the HFT at STAR. We have also reported on the measurements of the $J/\psi$ and $\Upsilon$ production in Au+Au collisions at $\sqrt{s_{\rm{NN}}}$ = 200 GeV enabled by the MTD. 


\end{document}